\documentclass[10pt,conference]{IEEEtran} 
\usepackage[switch,columnwise]{lineno}
\IEEEoverridecommandlockouts

\AtBeginDocument{%
  \providecommand\BibTeX{{%
    \normalfont B\kern-0.5em{\scshape i\kern-0.25em b}\kern-0.8em\TeX}}}

\newcommand{\myhash}{\raisebox{\depth}{\#}}

\usepackage{cite}
\usepackage{amsmath,amssymb,amsfonts}
\usepackage{algorithmic}
\usepackage{tabularx}
\usepackage{makecell}
\usepackage{tcolorbox}
\usepackage{multirow}
\usepackage{array}
\usepackage{booktabs}
\usepackage{graphicx}
\usepackage{textcomp}
\usepackage{xcolor}
\usepackage{xspace}
\usepackage[]{footmisc}
\usepackage{hyperref}
\usepackage{tikz}
\usepackage{colortbl}
\tikzset{%
pics/mycirc/.style args={#1}{
      code = {
\node [draw, circle,fill,outer sep=0pt, outer sep=0pt] {\color{white}#1};
}}}
\usepackage{tcolorbox}

\newcommand{\ignore}[1]{}
\newcommand{\mr}[1]{{\color{black} #1}}

\newcounter{finding}

\usepackage[flushleft]{threeparttable}
\usepackage{color}
\usepackage{makecell}
\definecolor{pblue}{rgb}{0.13,0.13,1}
\definecolor{pgreen}{rgb}{0,0.5,0}
\definecolor{pred}{rgb}{0.9,0,0}
\definecolor{pgrey}{rgb}{0.46,0.45,0.48}
\usepackage{multirow}

\usepackage{listings}
\def\BibTeX{{\rm B\kern-.05em{\sc i\kern-.025em b}\kern-.08em
T\kern-.1667em\lower.7ex\hbox{E}\kern-.125emX}}
\begin{document}

\newcommand{\technique}{TOGLL\xspace}

\title{
TOGLL: Correct and Strong Test Oracle Generation with LLMs}

\author{\IEEEauthorblockN{Soneya Binta Hossain}
\IEEEauthorblockA{\textit{Department of Computer Science} \\
\textit{University of Virginia}\\
sh7hv@virginia.edu}
\and
\IEEEauthorblockN{Matthew B. Dwyer}
\IEEEauthorblockA{\textit{Department of Computer Science} \\
\textit{University of Virginia}\\
matthewbdwyer@virginia.edu}}

\maketitle
\thispagestyle{plain}
\pagestyle{plain}
\begin{abstract}
Test oracles play a crucial role in software testing, enabling effective bug detection. Despite initial promise, neural methods for automated test oracle generation often result in a large number of false positives and weaker test oracles. While LLMs have shown impressive effectiveness in various software engineering tasks, including code generation, test case creation, and bug fixing, there remains a notable absence of large-scale studies exploring their effectiveness in test oracle generation. The question of whether LLMs can address the challenges in effective oracle generation is both compelling and requires thorough investigation.

In this research, we present the first comprehensive study to investigate the capabilities of LLMs in generating correct, diverse, and strong test oracles capable of effectively identifying a large number of unique bugs. To this end, we fine-tuned seven code LLMs using six distinct prompts on a large dataset consisting of 110 Java projects. Utilizing the most effective fine-tuned LLM and prompt pair, we introduce \technique, a novel LLM-based method for test oracle generation.  To investigate the generalizability of \technique, we conduct studies on 25 unseen large-scale Java projects. Besides assessing the correctness, we also assess the diversity and strength of the generated oracles. We compare the results against EvoSuite and the state-of-the-art neural method, TOGA. Our findings reveal that \technique can produce 3.8 times more correct assertion oracles and 4.9 times more exception oracles than TOGA. \mr{Regarding bug detection effectiveness, \technique can detect 1,023 unique mutants that EvoSuite cannot, which is ten times more than what TOGA can detect. Additionally, \technique significantly outperforms TOGA in detecting real bugs from the Defects4J dataset.}
\end{abstract}

\section{Introduction}
Software  dominates almost every aspect of our lives, including safety-critical domains such as healthcare and autonomous transportation. Even seemingly minor software bugs can cause large-scale system outages, security breaches, and loss of lives~\cite{ref1,ref2,porrello2012death}. Thus, ensuring software reliability through rigorous testing and bug detection is paramount. Test oracles, the foundation of effective testing, play a pivotal role in the early detection of software bugs \cite{hossain2023measuring,10.1145/3611643.3616265, schuler2013checked, zhang2015assertions}. 

Typically, a test suite consists of a set of test cases, where each test is composed of a \textit{test prefix} that exercises a certain part of a program, and a \textit{test oracle} verifies whether the executed behavior matches the expected behavior~\cite{myers2011art}. Test oracles can be categorized into two main types: 
\textit{assertion oracles}, which judge the correctness of the output state of the program, and
\textit{exception oracles}, which judge whether erroneous input states are detected by the SUT.

The efficacy of test oracles is determined by their ability to detect bugs.
For effective bug detection, test oracles should be \textit{correct} and \textit{strong}. A test oracle is considered correct if it aligns with the expected program behavior, thereby avoiding false alarms. Additionally, a test oracle is deemed strong if it can detect  deviations from intended program behavior. Just because a test oracle is correct does not mean it is strong, as demonstrated in ~\cite{10.1145/3611643.3616265}. 
\begin{figure*}[th]
    \centering
\includegraphics[width=0.9\linewidth]{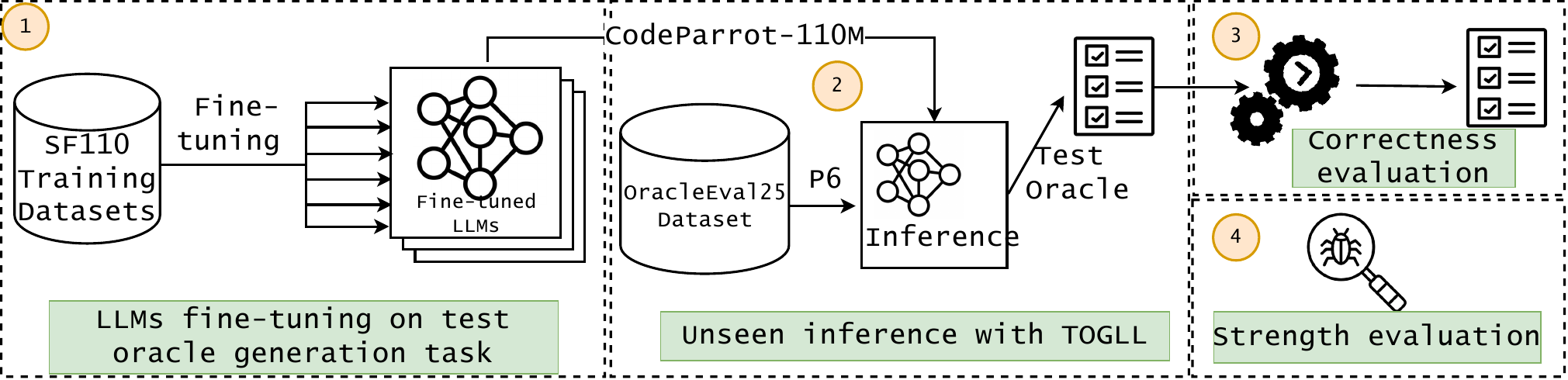}
	\caption{A general overview of our research approach to investigating LLM-based automated oracle generation. }
	\label{fig:overview}
\end{figure*}

Researchers have investigated the application of natural language processing (NLP) and pattern matching techniques for generating test oracles from code comments and natural language documentation~\cite{10.1145/3213846.3213872,10.1145/2931037.2931061,blasi2021memo,6227137,6200082}. 
These approaches can generate both assertion oracles, which verify that a program's actual output matches its expected output, and exception oracle, which capture the program's anticipated exceptional behaviors during testing.
Recent advancements have seen the application of neural techniques for generating test oracles, utilizing transformer-based models that learn from the method under test (MUT) and developer-written test cases~\cite{watson2020learning,tufano2022generating,tufano2020unit}. 
Building upon these methods, the development of TOGA~\cite{10.1145/3510003.3510141}—a neural-based test oracle generator—outperformed prior work in detecting real faults.  However, a recent study shows that this state-of-the-art neural-based test oracle generation method, when given a test prefix, the MUT and its docstring, generates assertion oracles only 38\% of the time.  Moreover, when it does generate assertion oracles 
47\% are false positives, and for exception oracles the false positive rate is 81\%. Despite significant research  focused on generating test oracles, automated generation of correct and strong test oracles has proven challenging and the state of the art suffers from: generalizability issues, high false positive rates and limited bug detection effectiveness~\cite{10.1145/3611643.3616265}. These challenges emphasize the ongoing need for further investigation, marking it as a significant and open research problem. 

Large language models (LLMs) have shown promise in automating diverse software engineering tasks such as code completion~\cite{li2023cctest}, program comprehension~\cite{nam2024using}, automated repair~\cite{jin2023inferfix}, and test generation~\cite{10329992,lemieux2023codamosa,siddiq2023exploring}.
Most prior work on ML-based test generation has focused on the generation of test prefixes -- inputs to the system under test --- with the goal of yielding substantial test coverage.  CodaMosa~\cite{lemieux2023codamosa} uses LLMs to escape coverage plateaus in search based test input generation and find it can improve coverage on small to moderate Python programs.  A recent attempt to use LLMs to generate test prefixes for large-scale Java programs was not very successful~\cite{siddiq2023exploring}; it achieved only 2\% statement coverage on large-scale Java projects compared to over 90\% for EvoSuite's~\cite{fraser2011evosuite} search-based approach.   The LLM-based TESTPILOT~\cite{10329992} approach generates both test prefixes and test oracles for JavaScript.  It produces better coverage than state-of-the-art search-based JavaScript test generation technique on a variety of small programs.  With regard to test oracles, TESTPILOT was shown to generate significant numbers of non-trivial assertions that
are distinct from those generated by other methods, but this work is limited to JavaScript and did not evaluate oracles correctness, diversity and strength aspects thoroughly.

In this paper, rather than task the LLM with generating both test prefix and test oracle, as TESTPILOT did, we decouple these problems.  We utilize state-of-the-art test generation methods, such as EvoSuite, to create test prefixes and concentrate on leveraging LLMs for test oracle generation. Initially, we fine-tune seven code LLMs, with sizes ranging from 110M to 2.7B parameters, employing six prompt formats that vary in the information they incorporate about the method under test (MUT), its documentation, and the test prefix. Employing the most effective model and prompt pair, we introduce our LLM-based method, \technique.

\mr{It is important to note that \technique generated oracles are not regression oracles. Regression oracles are generated based on the executed program behavior. For example, EvoSuite executes the code, collects all actual results values, knows the types, and then generates oracles based on those. This is why regression oracles are not capable of detecting bugs in current program versions. \technique, in contrast, does not execute the code and is not aware of the test execution results. It uses Javadoc documentation, MUT/MUT signature and test prefix to generate oracles. Some of the prompts (see Table \ref{tab:RQ1}) do not even use MUT (P1, P2) at all and some (P3, P4) only use MUT signature, not the entire implementation.} 

To evaluate \technique's generalizability on unseen data, we conduct a comprehensive study on 25 real-world large-scale Java projects and compare the correctness of the generated oracles with those produced by the SOTA neural method, TOGA~\cite{10.1145/3510003.3510141} and EvoSuite. \mr{Subsequently, we assess the strength of the generated oracles through a large-scale mutant detection study and a real bug detection study, highlighting the unique advantages of \technique-generated oracles over alternative methods. To the best of our knowledge, this represents the first extensive study to explore the application of LLMs in test oracle generation, examining seven code LLMs, six prompts, and three large datasets, while evaluating multiple aspects of the generated oracles, including correctness, diversity, and bug detection effectiveness.}

In summary, the main contributions of the paper lie in:
\begin{itemize}
\setlength\itemsep{1pt}
\item Constructing a large dataset from 110 large-scale Java projects to fine-tune code LLMs, aiming to investigate their performance in generating effective test oracles.
\item Evaluating the generalizability of fine-tuned LLMs in generating effective test oracles on 25 unseen projects and compare them with SOTA methods.
\item Evaluating the impact of different contextual information on oracle generation accuracy through six specially designed prompts.
\item Demonstrating that well-tuned and prompted LLMs can generate correct, strong, and diverse test oracles, with much lower false positive rates than SOTA-methods and can detect large number of unique bugs neither detected by EvoSuite nor SOTA neural method. 

\item Releasing all datasets, fine-tuned models, and code to enable replication of our study and the community to build on the work.
\end{itemize}
This work offers a promising starting point for LLM-based test oracle generation,
identifies directions for future improvement of such methods, and thereby has the potential to lead to impactful new capabilities for software engineers.

\section{Approach}

Figure \ref{fig:overview} presents an overview of our research approach. Step 1  of Figure \ref{fig:overview} depicts fine-tuning a collection of LLMs with different prompts for test oracle generation task; we discuss this in Section \ref{sec:fine-tune}. 
Step 2 of Figure \ref{fig:overview}selects the best performing fine-tuned model and prompt from Step 1 to define \technique and perform oracle inference on the unseen data.
Step 3 assesses the correctness of \technique generated oracles through test execution; we discuss this in Section 
\ref{test-unseen-data}. Finally, step 4 assesses the strength
of \technique-generated oracles using \mr{mutation testing and real bug detection from Defects4J}; we discuss this in Section \ref{mut-test} and \ref{defects4j}.

\subsection{Supervised Fine-tuning} \label{sec:fine-tune} This section covers the datasets, prompts, and LLMs that are fine-tuned for the test oracle generation task.

\subsubsection{SF110 Dataset}\label{training-data} We construct this dataset using the SF110 EvoSuite benchmark~\cite{10.1145/2685612}, which comprises 110 open-source Java projects with over 23,000 Java classes sourced from the SourceForge repository. This benchmark is widely recognized for unit test generation. To construct our dataset, we utilize the test cases generated by EvoSuite for each project. Given that a single test case can contain multiple assertion oracles or even a combination of assertion and exception oracles, we process each test case. This involve decomposing them into individual test cases, ensuring that each contains only a single assertion or exception oracle. For each decomposed test case, we extracted the method under test (MUT) along with its associated documentation to prepare the dataset. The dataset  is comprised of tuples $((p_i, m_i, d_i), o_i)$,
where, $p_i$ is the test prefix, $m_i$ is the MUT, and $d_i$ is the docstring (if available) and $o_i$ is the ground truth test oracle, either assertion or exception. This dataset is used to fine-tuning the LLMs and has been partitioned into three distinct subsets: 90\% for training, 5\% for validation, and another 5\% for testing.

\subsubsection{Prompt Designing}\label{prompts} 
Even though LLMs are powerful, they are not specifically designed and pre-trained
for test oracle generation. Thus, fine-tuning with effective prompts is necessary for optimal performance. We have designed six different prompts, each tailored to a specific use case scenario.

\begin{itemize}
\item Prompt 1 (\textbf{P1}) consists of only the test prefix (`prefix').   
\item Prompt 2 (\textbf{P2})  additionally includes the MUT documentation strings (`prefix + [sep] + doc') . The `[sep]' separator token is different based on different LLMs and their respective tokenizer. 
\item Prompt 3 (\textbf{P3}) instead of docstrings, includes the MUT signature (`prefix + [sep] + mutsig'). 
\item Prompt 4 (\textbf{P4})  includes both MUT signature and docstrings in addition to prefix (`prefix + [sep] + doc +[sep] + mutsig').
\item Prompt 5 (\textbf{P5}) includes the code for the entire MUT (`prefix + [sep] + mut').
\item Prompt 6 (\textbf{P6}) adds the docstring to P5 (`prefix + [sep] + doc + [sep] + mut').

\end{itemize}
\mr{These prompts are designed to cover diverse real use case scenarios. For example, when documents are not available, P1, P3, and P5 can be used. When the MUT implementation is not available,  P1 - P4 can be used, and when it is available,  P5 or P6 can be used. Designing these prompts allows us to assess their relative accuracies and the impact of different information on the oracle generation performance.}

\smallskip
\subsubsection{Code LLMs}\label{codemodels}

\mr{Decoder-only LLMs are most suitable for code generation tasks~\cite{hou2023large}. Smaller models are cost-effective, produce a lower carbon footprint~\cite{bender2021dangers}, and can be more easily fine-tuned with limited GPU resources. Therefore, in our study, we prefer decoder-only LLMs that are pre-trained on various code generation tasks within the software development life cycle (SDLC)\cite{hossain2024deep, hou2023large}, smaller in size, publicly available on Hugging Face\cite{huggingface2024}, and well-documented for fine-tuning. We chose seven LLMs because they are pre-trained on multiple programming languages while meeting all the above criteria.}

\smallskip

\noindent\textbf{CodeGPT~\cite{lu2021codexglue}}: It is a GPT-style language model with 110M parameters. We choose to fine-tune this model for two main reasons. Firstly, both our training and unseen inference datasets came from Java projects, and this model is also pre-trained on Java programming language. Secondly, this model is pre-trained on the code completion task, which aligns with our designed prompts. We fine-tune the microsoft/CodeGPT-small-java model from Hugging Face. 

\smallskip
\noindent\textbf{CodeParrot~\cite{codeparrot}}: In our study, we fine-tune the 110M parameters model, which was pre-trained for the code generation task on nine different languages: Java, JavaScript, PHP, Python, C\myhash, C++, GO, Ruby, and TypeScript. We have fine-tuned the codeparrot/codeparrot-small-multi model from Hugging Face. Our experimental study suggests that despite its small size, the fine-tuned CodeParrot model generalizes strongly on unseen data and can outperform larger models with billions of parameters.

\smallskip

\noindent\textbf{CodeGen~\cite{nijkamp2023codegen}}: CodeGen is a family of autoregressive language models with four different trainable parameter sizes: 350M, 2B, 6B, 16B. In this paper, we have fine-tuned 350M and 2B Multi variant models that are pre-trained on dataset that encompasses six programming languages: C, C++, Go, Java, JavaScript, and Python. Because of their great capability in comprehending and generating codes, we fine-tune them for test oracle synthesis task. 

\smallskip

\noindent\textbf{PolyCoder~\cite{xu2022systematic}}: PolyCoder is a family of large language models with three trainable parameter sizes: .4B, 2.7B, and 16B. This model was trained on 249 GB of code across 12 programming languages. In this paper we have fine-tuned .4B and 2.7B model for the test oracle generation task. 

\smallskip

\noindent\textbf{Phi-1~\cite{phi1model}}: Phi-1 is a transformer language model with 1.3 billion parameters, initially pre-trained on Python code from a variety of data sources for the task of Python code generation. Despite its primary training on Python, our study observed that it performs on par with other models when fine-tuned on Java code for generating test oracles.

\subsection{Unseen Inference}

\subsubsection{OracleEval25 Dataset}\label{test-unseen-data} To investigate the generalizability of the fine-tuned models, we utilize the dataset described in ~\cite{10.1145/3611643.3616265}, comprising 25 large-scale industrial standard Java projects. Of these, 17  originate from the Apache Commons Proper \cite{apache-commons-proper}, while the remainder are sourced from GitHub. These latter projects are characterized by large codebases with multiple modules, with up to 10k active users per project, and are frequently employed in software engineering research for the evaluation of test cases and test oracles ~\cite{just2014defects4j,schuler2013checked,zhang2015assertions}. Given their broad coverage of 21 domains and significant variation in developer metrics, the 25 artifacts selected for this study provide a robust framework for assessing fine-tuned model's ability to generalize across a spectrum of real-world projects. Overall, the dataset consists of 271K source lines of code (SLOC), 331K lines of JavaDoc, 1214K test suite SLOC and 223.5K test cases. Similar to the SF110 dataset, each input sample of this dataset contains the test prefix, MUT and docstring. We refer to this dataset as OracleEval25.

\subsubsection{\technique} From the fine-tuning results, we identify CodeGen-350M and CodeParrot-110M as the top two models due to their high accuracy. We also select the three best-performing prompts: P4, P5, and P6. \mr{As different fine-tuned models may generalize differently to unseen  data, and due to the differences between datasets, different prompts may work better. For this reason, we conduct a small-scale study, performing unseen inference with these two models and three prompts on projects from the OracleEval25 dataset.} Our results suggest that even though CodeGen-350M is three times larger than CodeParrot-110M, CodeParrot-110M fine-tuned with P6 performs better on the unseen project than CodeGen on several metrics: total input processed, compilation error, and false positive count. Furthermore, the inference time with CodeGen-350M is longer than CodeParrot. Consequently, we define \technique as a fine-tuned CodeParrot-110M operating on P6. We perform rest of the study with \technique on the unseen  OracleEval25 dataset \ref{test-unseen-data}. 

\subsubsection{Baseline} \label{toga} 
\begin{figure}[t]
    \small\centering
\includegraphics[width=.7\columnwidth]{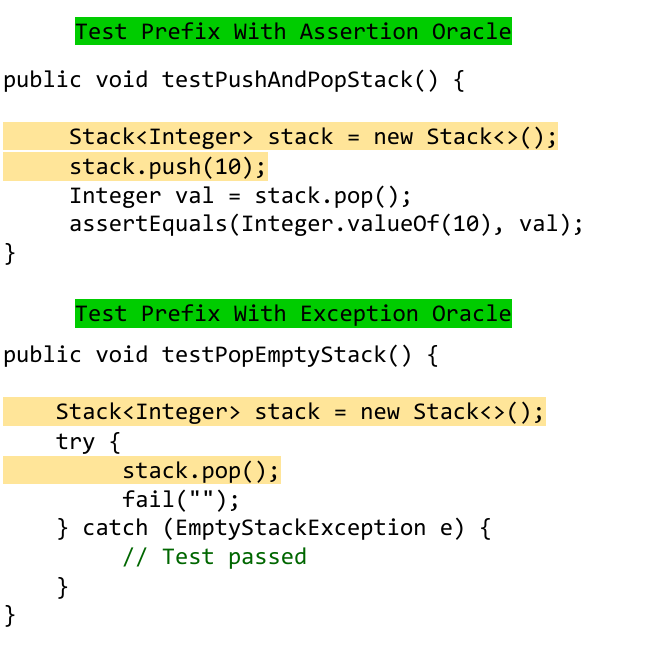}
	\caption{Test cases with assertion and exception oracles. The test prefix part is marked with yellow color.}
	\label{fig:example}

\end{figure}

TOGA is a neural method for both assertion and exception oracle generation \cite{10.1145/3510003.3510141}. TOGA utilizes CodeBERT \cite{feng2020codebert} model
as backbone and fine-tuned for the oracle generation task. TOGA takes the test prefix generated by EvoSuite, its associated MUT and docstring (if available) and generates either an assertion or an exception oracle. An example of EvoSuite generated test cases with assertion and exception oracle is shown in Figure \ref{fig:example}. The prefix part is marked with yellow color. The first test prefix involves pushing the number 10 onto the stack, followed by popping this value from the stack. For this scenario, an assertion oracle is anticipated to validate that the value retrieved is indeed 10. The second test attempts to pop from an empty stack, expecting the MUT to throw an exception. If no exception is thrown, the test should fail.

We selected TOGA as our baseline method for several reasons: 1) TOGA represents the SOTA neural method, having surpassed specification, search, and neural-based techniques by detecting 57 bugs in Defects4J~\cite{just2014defects4j}; 2) TOGA utilizes the same set of information as P6, which is our most comprehensive prompt, thereby providing a fair basis for comparison; 3) TOGA has been previously evaluated on the OracleEval25 dataset, which is also employed in our study, ensuring consistency in our comparative analysis~~\cite{10.1145/3611643.3616265}.

\subsubsection{Metrics} With \technique, we generate test oracles for all projects from the OracleEval25 dataset. This dataset framework includes all necessary artifacts for executing the generated test oracles when integrated with the respective test prefixes. Consequently, we integrate the generated oracles into the test cases and execute the complete test suites. We calculate accuracy as the success rate, which denotes the percentage of non-empty test oracles that were successfully passed during test execution, indicating their alignment with expected program behavior. We compute the success rate using Equation \ref{eq:1}
\begin{equation}\label{eq:1}
\textrm{SuccessRate}(P) = \frac{T - (T_{ce} + T_{fp} + T_{em})}{T}
\end{equation}
where $T$ is the number of total test prefixes in project $P$, 
$T_{ce}$ is the number of compilation errors, 
$T_{fp}$ is the number false positive tests, and 
$T_{em}$ the number of empty oracles. We compute the SuccessRate metric for both assertion and exception oracles predicted by \technique and compare them to those of our baseline, TOGA.

\mr{To assess the bug detection effectiveness of the \technique-generated oracles, we perform both mutation testing and real bug detection study on Defects4J. Mutation testing introduces small changes, or mutations, to the source code to create numerous slightly altered versions of the software, called \textit{mutants} \cite{coles2016pit, petrovic2021practical}. Studies showed that there is a statistically significant
correlation between mutant detection and real fault detection~\cite{10.1145/2635868.2635929,andrews2005mutation,petrovic2021does}. A test suite's ability to detect these mutants can be a direct measure of its bug detection effectiveness \cite{zhang2015assertions,schuler2013checked}. Therefore, we compute the mutant killing score as a metric to assess and compare the bug detection effectiveness of \technique-generated oracles with both EvoSuite and TOGA. Additionally, we also record unique mutants killing score which indicates the unique capability of each oracle generation method. Besides the mutation study, we also perform real bug detection study on the Defects4J dataset and compare the results with the TOGA baseline. }
\begin{table*}[ht]
\small\centering
\caption{Test oracle generation performance of LLMs on different prompts. Top two accuracies for each prompt are shown in \textbf{bold}. The last row shows prompt-wise average.}\label{tab:RQ1}

\resizebox{\linewidth}{!}{%
\begin{tabular}[t]{c|c|c|c|c|c|cccc}
 \cmidrule{1-7} \cmidrule{10-10}
  \textbf{Code LLM} & \textbf{\thead{P1}} & \textbf{\thead{P2}} &\textbf{\thead{P3}} &\textbf{\thead{P4}} &\textbf{\thead{P5}} &\textbf{\thead{P6}}& & & \textbf{Prompt Details}
 \\
  
\cmidrule{1-7} \cmidrule{10-10}
\textbf{CodeGPT-110M} & 54.4  &
63.5 &
72.7 &
73 &
75 &
74.1 & & & \textbf{P1}: prefix
\\\cmidrule{1-7} \cmidrule{10-10}

\textbf{CodeParrot-110M} & \textbf{56.4} &
\textbf{65.6} &
\textbf{76.1} &
\textbf{76.2} &
\textbf{77.7} &
\textbf{77.7} & & & \textbf{P2}: prefix + [sep] + doc.
\\\cmidrule{1-7} \cmidrule{10-10}

\textbf{CodeGen-350M} & \textbf{56.7} &
\textbf{66.3} &
\textbf{77.4} &
\textbf{77.5} &
\textbf{79.3} &
\textbf{79} & & &\textbf{P3}: prefix + [sep] + mutsig.
\\\cmidrule{1-7} \cmidrule{10-10}

\textbf{PolyCoder-.4B} & 56.4 &
64.3 &
74.5 &
74.9 &
76.8 &
76.8 & & & \textbf{P4}: prefix + [sep] + doc. + [sep] + mutsig
\\\cmidrule{1-7} \cmidrule{10-10}

\textbf{Phi-1(1.3B)} & 53.42 &
61.5 &
75.5 &
73.4 &
77.6 &
74.3 & & & \textbf{P5}: prefix + [sep] + mut
\\\cmidrule{1-7} \cmidrule{10-10}

\textbf{CodeGen-2B} & 55.7 &
65.5 &
75 &
74.7 &
76.5 &
76.8 & & &\textbf{P6}: prefix + [sep] + doc. + [sep] + mut
\\\cmidrule{1-7} \cmidrule{10-10}

\textbf{PolyCoder-2.7B} &
55 &
64 &
74.8 &
74.7 &
76.6 &
76.9  & & &\\\cmidrule{1-7}

\textbf{Avg:} &
55.4 & 64.4 & 75 & 75 & 77 & 76.5 & & &\\\cmidrule{1-7}

\end{tabular}
}
\end{table*}

\section{Experimental Study}
To assess the efficacy of LLM-generated test oracles, this study investigates several key areas: effectiveness of different code LLMs in generating test oracles, the impact of different prompts on the oracle generation accuracy, the ability of fine-tuned models to generalize to unseen data, comparative accuracy among different oracle types, diversity of the generated oracles and the overall efficacy of these oracles in detecting bugs. To this end, we answer the following five research questions.

\smallskip

\textbf{RQ1:} What LLM and prompting approaches are effective for test oracle generation? \textbf{\textit{Increasing context in the prompt improves accuracy of oracle generation up to 79\% for a range of LLMs, but smaller LLMs meet or exceed the performance of larger LLMs when fine tuned.}}

\smallskip

\textbf{RQ2:} How well does \technique, i.e., fine-tuned model generates correct test oracles for unseen projects? \textbf{\textit{\technique generates up to 4.9 times the number of correct test oracles compared to prior neural based approach.}}

\smallskip

\textbf{RQ3:} How diverse are LLM-generated assertions relative to those generated by EvoSuite? \textbf{\textit{\technique-generated oracles vary substantially in terms of the specific assert statements used as well as in the variables and expressions references in those oracles compared to those generated by EvoSuite.}}

\smallskip

\textbf{RQ4:} How strong are TOGLL-generated assertions at identifying unique mutants? \textbf{\textit{\technique generated oracles kill nearly 10 times more unique mutants than prior SOTA method and fall only 17\% short of EvoSuite's bug detection effectiveness.}}

\smallskip

\mr{\textbf{RQ5:} How strong are TOGLL-generated oracles at detecting real bugs from Defects4J? \textbf{\textit{\technique-generated exception and explicit assertion oracles detected 106\% more bugs compared to the prior SOTA method TOGA.}}}

\subsection{RQ1: Selecting LLMs and Prompts for Oracle Generation}
In this research question, we fine-tune various large language models, as outlined in Section \ref{codemodels}, utilizing six unique prompts discussed in Section \ref{prompts}. The purpose of this study is to determine the most effective combination of model and prompt format for the test oracle generation task. This approach allows us to understand the impact of prompts on model efficacy and guides the selection of the most effective model-prompt pair for a deeper investigation. 

\subsubsection{Experimental Setup} For this study, we utilized the SF110 training dataset discussed in Section~\ref{training-data}, which includes 159,073 input samples. Each sample contains a unique identifier, the method under test (MUT), a test prefix, and a docstring, from which we constructed six prompts as outlined in Section \ref{prompts}. For the fine-tuning process of the code LLMs, we employed a GPU setup consisting of 4 A100 GPUs, each with 40GB of memory. To facilitate parallel processing and accommodate larger models such as CodeGen-2B and PolyCoder-2.7B within the 40GB memory limit, we utilized PyTorch Accelerator and DeepSpeed~\cite{rasley2020deepspeed}. The models were fine-tuned using a learning rate of 2.5e$^{-5}$ and the AdamW optimizer. Each LLM was fine-tuned across all six prompts, and accuracy was computed on the validation dataset using the `exact match' metric. This metric may underestimate the overall accuracy, as a generated oracle may not be an exact match but can still be correct, as illustrated in Figure ~\ref{fig:rq3}.

\subsubsection{Results} Table \ref{tab:RQ1} displays the specifics of each prompt—detailing the information used in each prompt—on the right side. P1 utilizes only the test prefix as input, prompting the model to generate either an assertion or an exception oracle. P2 adds a docstring in addition to the test prefix, while P3 includes a method signature. P4 includes both docstring and method signature after the test prefix. P5 includes the entire MUT code with the test prefix, whereas,  P6 contains both the docstring and entire MUT code with the test prefix. On the left side of Table \ref{tab:RQ1}, we present the performance of seven fine-tuned code models in generating test oracles for all six prompts.

The average accuracy of the fine-tuned models on P1 is around 55.4\%. If we add the available docstring in the prompt, the accuracy is increased by around 10\%. Instead of adding the docstring, if we add the method signature to the prompt, the accuracy is increased by 20\%, resulting in on average 75\% accuracy for P3. P4 includes both docstring and method signature with test prefix, which only see a marginal improvement for few models. Most accuracy improvement was observed when we add the entire code of the MUT in P5 and P6, with P5 achieving highest accuracy of 77\%. We also observe that adding more context can also decrease the accuracy, such as for P5 and P6. Overall, we see a consistent accuracy improvement as we go from prompt P1 to P5. 

We observe that including method signature in the prompt works significantly better than including the docstring. Even when including both method signature and docstring -- P4 -- only marginal improvement was observed. This observation also true for P5 and P6, adding the docstring did not improve much accuracy when either method signature or entire method code is available. However, from prompt P1 to P2, accuracy is improved by 10\%, indicating the value of docstrings. The quality of the docstrings is not considered in this study. This is a promising future direction as we believe that documentation that combines natural language description of requirements with references to API elements might yield better results.

Among the seven code models evaluated, CodeGen-350M emerged as the top performer, closely followed by CodeParrot-110M, both highlighted in bold. The three top performing prompts are P4, P5, P6. \mr{RQ2 is performed on the unseen OracleEval25 dataset. Our investigation shows that CodeParrot-110M generalizes better on the unseen data than CodeGen-350M. Also, although, P5 works better in this study we find that P6 works better on the 
OracleEval25 dataset, which has more Javadoc (for 75\% samples) compared to the SF110 dataset (45\%) and P6 can leverage those Javadoc to generate better test oracles.}

\begin{tcolorbox}

\textbf{RQ1 Findings:} 
LLMs are capable of accuracy levels above 79\% in generating test oracles.  Prompts containing increased contextual information enable higher accuracy, but perhaps surprisingly larger LLMs when fine-tuned do not offer significant advantages.
\end{tcolorbox}
\begin{figure*}[ht]
    \small\centering
\includegraphics[width=.9\linewidth]{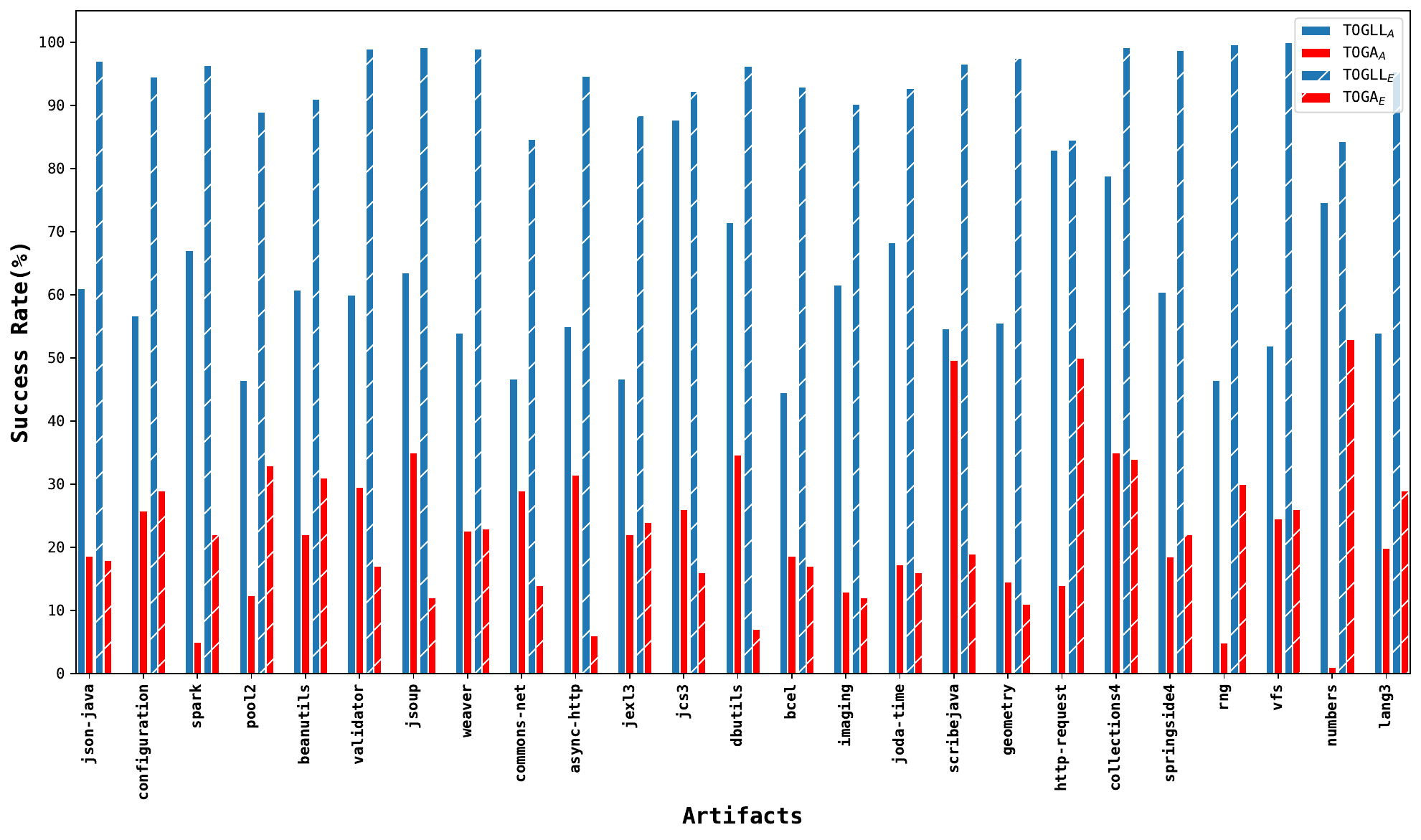}
	\caption{Comparison of correct test oracle generation performance of \technique vs. TOGA}
	\label{fig:rq2}
\end{figure*}

\subsection{RQ2: Assessing Test Oracles Correctness}\label{sec:rq2}
RQ1 investigates various LLMs' test oracle generation performance when they are fine-tuned on the training 
portion of the SF110 dataset and 
their accuracy is evaluated on the validation portion of the same dataset. However, a fine-tuned model may not generalize when applied to unseen data, a common challenge in LLM applicability~\cite{hou2024large}. To investigate the generalizability of our method \technique, we perform this study on the unseen OracleEval25 dataset discussed in Section ~\ref{test-unseen-data}. 
A test oracle is correct if it captures the expected behavior of a software system.
Generated test oracles that do not compile, that disagree with software behavior, or that are vacuous are considered
incorrect.  We study the rate of success in generating correct oracles for both \technique and TOGA.

\subsubsection{Experimental Setup} We generate test oracles for the OracleEval25  unseen dataset using our method \technique, which is a fine-tuned CodeParrot-110M model operating on P6, which includes the MUT, test prefix and docstrings. This dataset consist of a total of 202,434 assertion prefixes requiring \technique to generate an assertion oracle for each input sample and 21,123 exception prefixes requiring \technique to generate an exception oracle. We have used a single A100 GPU with 10GB of memory to perform the inference task. The dataset keeps a record of the metadata so that the generated test oracles can be integrated with the test prefixes, enabling test validation to be performed later. 

Unlike RQ1, which assessed accuracy using the `exact match' metric, RQ2 involves executing the integrated test cases—combining the test prefix with the generated test oracles—to collect detailed metrics for each type of oracle. For both assertion and exception oracles, we compute success rates across all 25 projects using Equation ~\ref{eq:1}, and compare these rates to those achieved by TOGA. \mr{Counting empty and non-compiling assertions is straightforward. For false positives, a failing test oracle can indicate either an incorrect oracle or a bug in the MUT. If the MUT is known to be correct, a failing test oracle signals a false alarm. We use this 
well-established technique for test oracle evaluation \cite{10.1145/3611643.3616265}
to evaluate TOGLL-generated oracles. The OracleEval25 dataset includes the latest stable releases of projects with no known bugs. Therefore, any failure of a generated test oracle suggests it is incorrect.}

\ignore{
Success rate represents the percentage of non-empty test oracles that passed during test execution, indicating that they align with expected program behavior. We compute the success rate using Equation \ref{eq:1}
\begin{equation}\label{eq:1}
\textrm{Success Rate}(A) = \frac{T - (T_{ce} + T_{fp} + T_{em})}{T}
\end{equation}
where $T$ is the number of test prefixes in the artifact $A$, 
$T_{ce}$ is the number of compilation errors, 
$T_{fp}$ is the number false positive tests, and 
$T_{em}$ the number of empty oracles.}

\subsubsection{Results}
In Figure \ref{fig:rq2}, we present and compare our LLM-based method, \technique's correct test oracle generation performance with deep-learning-based method, TOGA.
The Y-axis displays success rate as a percentage. The X-axis enumerates the names of 25 projects. Each project is represented by two sets of vertical bars: blue for \technique and red for TOGA. Solid blue bars show \technique's success rate in generating assertion oracles, while blue bars with diagonal lines illustrate its success in generating exception oracles. Similarly, solid red bars represent TOGA's assertion oracle success rates, and red bars with diagonal lines show its performance in generating exception oracles.

Figure \ref{fig:rq2} shows that \technique achieved significantly higher success rate across all projects for both assertion and exception oracle generation. For example, for apache commons-number, TOGA could achieve a success rate of only 1\%, whereas, \technique achieve significantly higher accuracy of 74.7\%. Similarly, for http-request, TOGA achieved 14\% where as \technique achieved 83\% success rate. Statistical analysis reveals a significant difference in the average success rates, with \technique achieving a mean of 63\% compared to TOGA's mean of 16.4\%. A t-test yielded a t-statistic of 12 and a p-value of $1.1e^{-11}$, indicating a statistically significant difference in performance between the two methods, with \technique outperforming TOGA in terms of correct test oracle generation. 

\mr{TOGA uses a non-ML method to generate five types of assertion candidates using predefined templates, and then an ML model ranks the candidates to find the best assertion. We found that around 62\% of the time, the model could not generate any assertions. Additionally, TOGA uses the most frequent constants and variables to generate assertEquals oracles, resulting in high false positives. TOGLL is not restricted to certain types of assertions and does not use frequently appearing values to generate assertions. Moreover, TOGLL makes more effective use of Javadoc than TOGA, which showed almost no improvement with Javadoc. TOGLL's training data and model backbone could also contribute to its superior performance, but this requires further investigation.}

\mr{For exception oracle, the performance difference is even higher. TOGA's average success rate is 18.86\%, while \technique achieves a significantly higher average of 93.4\%. A Mann-Whitney U test further confirms a statistically significant performance difference between \technique and TOGA, highlighting \technique's superior performance in generating exception oracles compared to TOGA (U = 625.0, p = $1.4e^{-09}$). Javadoc typically specifies exception conditions for a MUT and \technique's ability to leverage these Javadoc and the specified exception conditions is a key factor contributing to its superior performance.}

\begin{tcolorbox}

\textbf{RQ2 Findings:} 
\technique generates significantly more correct test oracles than TOGA; bettering it by 3.8 times and 4.9 times for assertion oracles and exception oracles, respectively.
\end{tcolorbox}

\subsection{RQ3: Investigating Diversity in Oracle Generation}
\mr{Since \technique is fine-tuned on EvoSuite-generated test oracles, one concern is whether \technique simply learns to generate EvoSuite oracles. To this end, RQ3 investigates the diversity of the generated test assertions.} More specifically, we investigate whether during  unseen inference \technique follows the assertion distribution observed during training and also to what extent the generated assertions exactly match the ground truth assertions. These metrics can provide insights into the diversity of the generated assertions. Diversity is an important aspect of test oracle generation, allowing \technique to complement developer written oracles or those generated by other tools and thereby enable detection of faults that would  otherwise be undetected.
In RQ4 and RQ5, we investigate the ability of \technique to detect unique faults.

\subsubsection{Experimental Setup} To set up this study, we utilize the training dataset and results from unseen inference in RQ2. We divided the assertions into seven categories, consisting of the most frequently used JUnit assertions investigated in previous studies~\cite{zhang2015assertions, 10.1145/3510003.3510141, 10.1145/3611643.3616265}. We count the total number of assertions in each category and calculate the distribution relative to the total number of assertions in the training data. We follow a similar process to compute the categorical distribution of \technique generated assertions during unseen inference. Furthermore, we compute the percentage of \technique generated assertions that are an exact match for ground truth assertions in our dataset.

\subsubsection{Results} Table \ref{tab:rq3}, Column 1 lists the assertion categories, Column 2 shows the distribution of different types of assertions within the training dataset. Column 3 presents the breakdown of the oracles generated by \technique for the unseen dataset. Column 4 presents the percentage of exact match assertions for each category, reflecting the instances where the \technique-generated assertions exactly match the ground truth assertions.

There is a clear evidence that the distribution of assertions generated by \technique is quite different from the training data. The training data has a sharp peak for \texttt{assertEquals}, whereas the inference data has lower peaks for two categories. Moreover the tails of the distributions have a much higher share of samples indicating that generation of assertions is much less concentrated than training data. With regard to assertion category \technique is more diverse in generating assertion oracles.

\begin{figure}[t]
    \small\centering
\includegraphics[width=.99\columnwidth]{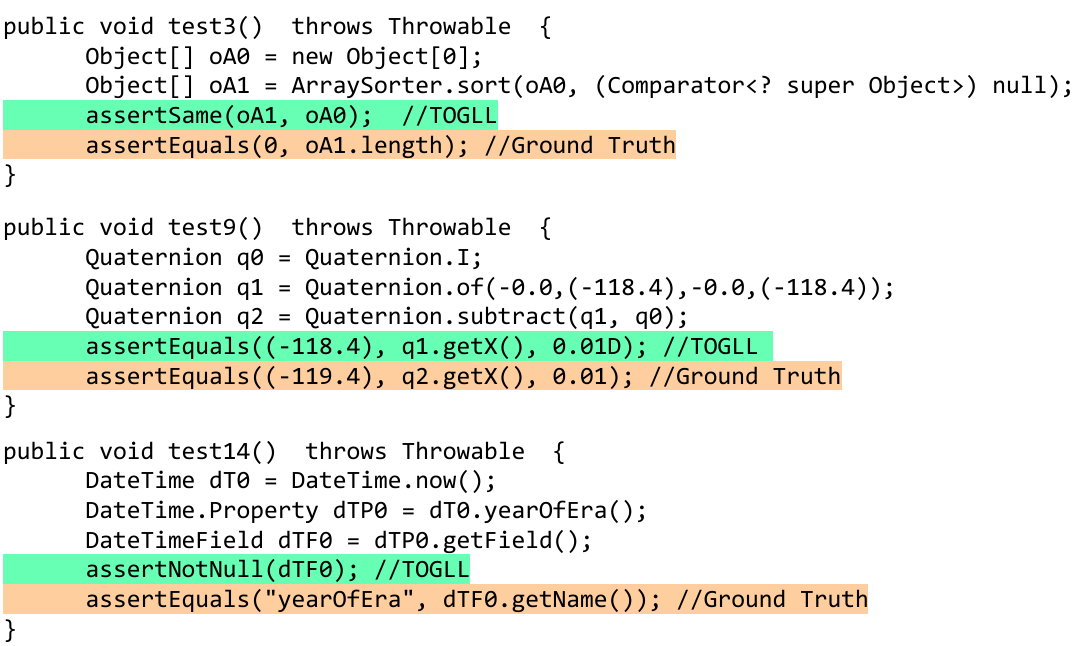}
	\caption{Diverse yet correct test oracles generated by \technique}
	\label{fig:rq3}
\end{figure}

We see that out of the 194k generated assertions, only 18k are exact matches, indicating that the generated assertion are diverse relative to provided assertions in the dataset.
We observe this diversity can come multiple ways.
For example, the generated assertion may come from a different category, like the first and third examples in Figure~\ref{fig:rq3}, or may vary the assertion arguments, like the second example in Figure~\ref{fig:rq3}.
We observe that ground truth oracles do not check all variables within the prefix which gives \technique an opportunity to generate diverse assertion oracles. 
We see this in the first example where the \technique generated assertion is 
requiring that sort yield the same array, whereas the EvoSuite assertions just
checks the length of the array.  Similarly the second example shows a \technique
generated assertion for the \texttt{q1} variable, whereas EvoSuite's assertion targets
the \texttt{q2} variable. For the third example, \technique generates a different type of oracle. Not only are these \technique-generated assertions diverse, but they potentially add unique bug detection power which we explore in the next RQs. \mr{We can conclude that fine-tuning with EvoSuite oracles does not constrain \technique to EvoSuite-like oracles, as it generates distinct yet correct assertions involving various assertion types, variables, and expressions, with only a 9.5\% exact match rate. In RQ4, we demonstrate that these distinct oracles detect 1,023 unique bugs that EvoSuite could not detect (Table \ref{tab:rq4}). Pre-trained on a vast codebase, \technique's LLM backbone has likely encountered both developer-written and automated assertions. By leveraging this pretraining knowledge, \technique can generate diverse yet accurate assertions with strong bug detection effectiveness.}

\begin{table}[t]
\small\centering
\caption{Distribution of assertion oracles across different categories}\label{tab:rq3}
\resizebox{\linewidth}{!}{

\begin{tabular}[t]{l|c|c|c}

\toprule
  \multirow{2}{*}{\thead{\textbf{Assertion} \\\textbf{Category}}} & \multicolumn{3}{c}{\textbf{Assertion Distribution}} \\\cline{2-4}
  & \textbf{\thead{Training}} & \textbf{\thead{Inference}} & \textbf{\thead{Exact Match}}\\
  \midrule

assertNotNull &  
18,033 \textbf{(16.1\%)}& 

79,476 \textbf{(40.8\%)}& 

9203 \textbf{(11.6\%)} \\\midrule

assertEquals & 
81,766  \textbf{(72.8\%)}& 
79,680  \textbf{(40.9\%)}& 7217 \textbf{(9\%)}
\\\midrule

assertNull&
10,484 \textbf{(9.3\%)}& 
6,542 \textbf{(3.4\%)}&  1660 \textbf{(25.3\%)}\\\midrule

assertSame & 
636 \textbf{(0.6\%)}& 
5,791 \textbf{(3\%)}&  406 \textbf{(7\%)}\\\midrule

assertFalse & 
651  \textbf{(0.6\%)}& 
12,575   \textbf{(6.5\%)}&  43 \textbf{(0.3\%)}  \\\midrule

assertNotSame  & 
334   \textbf{(0.3\%)}& 
3,210 \textbf{(1.6\%)}&  89 \textbf{(2.8\%)}\\\midrule

assertTrue  & 
349  \textbf{(0.3\%)} & 
3,268 \textbf{(1.7\%)} &  12 \textbf{(0.37\%)} \\\midrule

syntx. incorrect & 
-   & 4,329 \textbf{(2.2\%)}
 &  - \\\midrule

\textbf{Total:}  & \textbf{112,253} & \textbf{194,871}  &  \textbf{18,630} \\

\bottomrule
\end{tabular}
}
\end{table}
\raggedbottom

\begin{tcolorbox}
\textbf{RQ3 Finding:} 
\technique assertions are diverse with respect to both the assertion statement used and the variables and expressions targeted for observation in those assertions. 
\end{tcolorbox}

\subsection{RQ4: Assessing Mutant Detection Effectiveness}\label{mut-test}

\begin{table}[ht]
\small\centering
\caption{Mutant detection performance of \technique, EvoSuite and TOGA generated assertions.}\label{tab:rq4}
\resizebox{.8\columnwidth}{!}{%

\begin{tabular}[t]{l|c|c|c|c}
  \toprule
\multirow{2}{*}  \textbf{Artifact} & \multicolumn{4}{c}{\textbf{Bug Detected by}} \\
\cline{2-5}
& \textbf{\thead{EvoSuite\\ Assertion}} & \textbf{\thead{EvoSuite \\Unique}} & \textbf{\thead{\technique \\Assertion}} & \textbf{\thead{\technique \\Unique}}\\
\midrule
   
\textbf{http} &  64 &  38 &  29 &  3\\\hline
\textbf{json} &  328 &  110 &  249 &  31\\\hline
\textbf{beanutils} &  551 &  46 &  516 &  11\\\hline
\textbf{collections4} &  248 &  104 &  162 &  18\\\hline
\textbf{dbutils} &  108 &  14 &  98 &  4\\\hline
\textbf{jsoup} &  598 &  177 &  509 &  88\\\hline
\textbf{imaging} &  1,869 &  629 &  1,315 &  75\\\hline
\textbf{lang3} &  2,301 &  397 &  2,063 &  159\\\hline
\textbf{configuration} &  271 &  51 &  266 &  46\\\hline
\textbf{jexl3} &  697 &  73 &  638 &  14\\\hline
\textbf{joda-time} &  1,545 &  446 &  1,323 &  224\\\hline
\textbf{net} &  747 &  175 &  592 &  20\\\hline
\textbf{pool2} &  155 &  8 &  150 &  3\\\hline
\textbf{spark} &  407 &  63 &  355 &  11\\\hline
\textbf{validator}  &  602 &  52 &  560 &  10\\\hline
\textbf{scribejava} &  172 &  31 &  143 &  2\\\hline
\textbf{bcel} &  1,122 &  467 &  698 &  43\\\hline
\textbf{numbers} &  871 &  168 &  726 &  23\\\hline
\textbf{springside4} &  1,160 &  243 &  938 &  21\\\hline
\textbf{vfs2} &  339 &  37 &  310 &  8\\\hline
\textbf{rng} &  554 &  158 &  527 &  131\\\hline
\textbf{jcs3}  &  621 &  81 &  548 &  8\\\hline
\textbf{async-http} &  69 &  10 &  59 &  0\\\hline
\textbf{weaver} &  20 &  2 &  18 &  0\\
  
\hline
\textbf{Total:} &\textbf{15,985}	& \textbf{3,690}	& \textbf{13,318}	&\textbf{1,023} \\	
\hline
&	& \textbf{TOGA:} & 	\textbf{6,893}  &\textbf{105}\\			

\bottomrule
\hline
\end{tabular}
}
\end{table}

Correct test oracles are essential for aligning with program requirements; however, correctness alone may not suffice to detect a diverse range of bugs. The strength of test oracles is equally crucial. In RQ4, we investigate the mutant detection effectiveness of \technique-generated assertions and compare them with the SOTA EvoSuite and TOGA.

\subsubsection{Experimental Setup} To set up this experiment, we generate three distinct test suites (T$_{\technique}$, T$_{ES}$, T$_{IO}$) for each of the 25 projects from the OracleEval25 dataset. T$_{\technique}$ comprises test cases with \technique-generated assertions, excluding all test cases with non-compiling and incorrect (i.e., false positive) assertions produced by \technique. We recorded the set of final test cases, including the test class and the unique test identifier. For constructing T$_{ES}$, we use the same set of tests, but the test assertions were generated by EvoSuite. For T$_{IO}$, the same set of test cases was retained; however, this time, the test cases did not contain any test assertions. The three test suites share identical counts of test cases, test case prefixes, and assertion totals. The unique aspect distinguishing them is the method used for generating assertions.

We utilize mutation testing to introduce a wide array of bugs into the source code of 25 projects. Unlike bug benchmarks such as Defects4J ~\cite{just2014defects4j} or QuixBugs~\cite{10.1145/3135932.3135941}, which are limited in both the number and diversity of bugs, mutation test injects thousands of diverse bugs. 
We employ PIT~\cite{coles2016pit}, a state-of-the-art mutation testing tool for Java programs. PIT offers several advantages over other tools, as it generates meaningful mutants and fewer equivalent mutants~\cite{kintis2018effective,7927997,7359265}. We have used PIT-1.9.8 to generate a total of 69,793 \texttt{strong} mutants. For each mutated project, we run all three test suites to detect mutants. Mutants detected by T$_{IO}$ are those that can be detected through implicit assertions, meaning test prefixes alone suffice for their detection without the need for any explicit assertions. 
Computing T$_{IO}$ allows our study to follow the recommendations of \cite{10.1145/3611643.3616265} by removing mutants killed by implicit oracles from study results. We calculate the total number of mutants detected by EvoSuite, \technique and TOGA assertions as well as the unique mutants killed by assertions generated by each method.

\subsubsection{Results} In Table \ref{tab:rq4}, Column 1 shows the names of 25 projects. Columns 2 and 3 present the total and unique mutants detected by EvoSuite-generated assertions, respectively. Columns 4 and 5 present the total and unique mutants detected by \technique assertions, respectively. Row 27 shows the total mutants detected across all projects by each method, whereas row 28 presents the performance of TOGA, in detecting total and unique mutants.

Out of the 69,793 mutants generated and covered by test cases, EvoSuite detected a total of 15,985 mutants, including 3,690 unique mutants not detected by \technique assertions. Conversely, \technique assertions identified 13,318 mutants, with 1,023 being unique mutants. 

For all 25 projects, test cases were generated by EvoSuite, as discussed in Section \ref{test-unseen-data}. EvoSuite, a search-based dynamic approach, generates assertions based on executed program behavior and actual result values. Thus, it is expected to produce strong assertions capable of detecting a large number of mutants. \technique, in contrast, is a static approach that utilizes the test prefixes generated by EvoSuite, but lacks knowledge of any values computed during test execution. Despite this disadvantage, \technique-generated assertions detected 83\% of the EvoSuite-detected mutants.
This significantly outperformed TOGA, which detected only 43\% of the EvoSuite-detected mutants under the same conditions.  

\begin{figure}[ht]
    \small\centering
\includegraphics[width=.9\columnwidth]{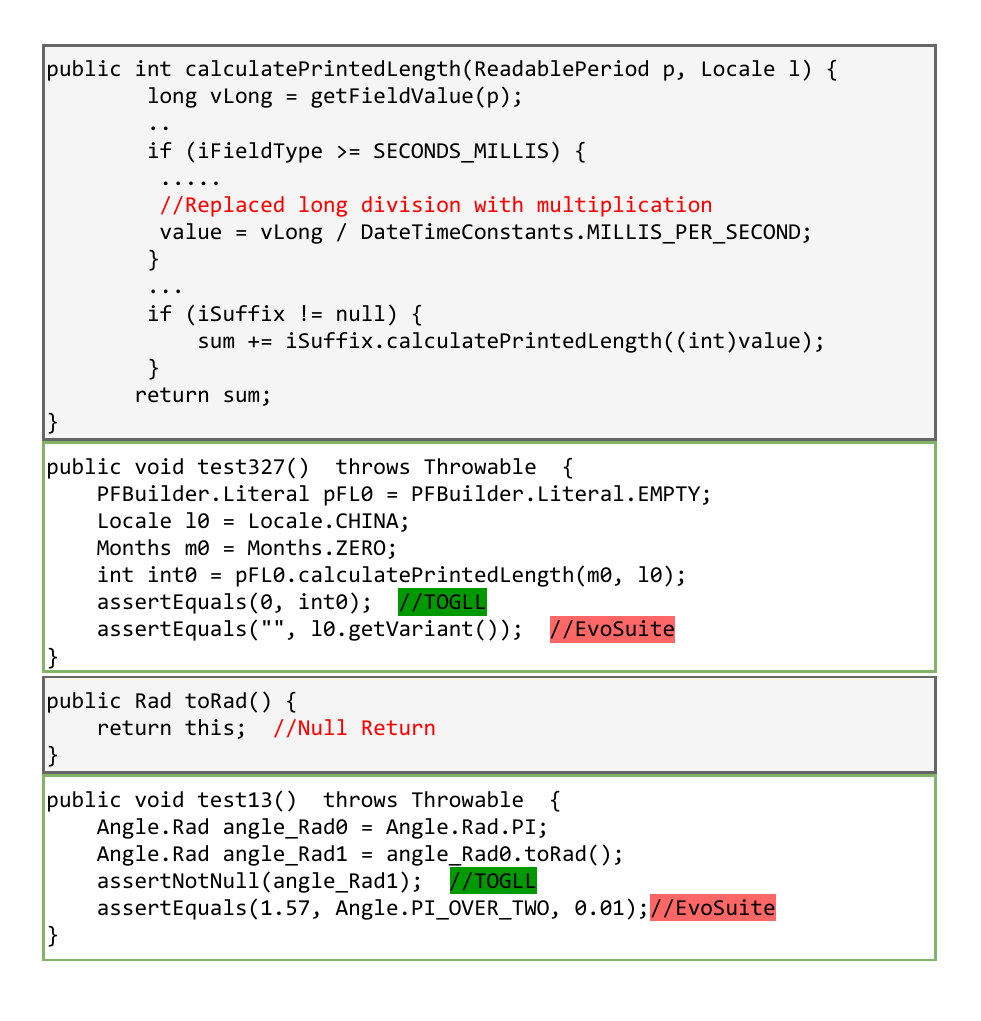}
	\caption{\technique generated assertions detecting unique mutants that EvoSuite generated assertions  mcan not.}
	\label{fig:rq4}
\end{figure}

\technique generated assertions were able to detect 1,023 mutants that EvoSuite missed, using the same set of test prefixes. This is also a significant improvement over TOGA, which only identified 105 unique mutants. In total, \technique killed nearly twice the number of mutants as TOGA, but when viewed relative to TOGA
as a baseline \technique killed 9.7 times the number of unique mutants.
This demonstrates that \technique can generate a significant number of strong test assertions and highlights \technique's capability to complement EvoSuite's assertions.

We perform a qualitative study of the \technique assertions that could kill unique mutants.
This study was opportunistic in that we look into the 1023 unique mutants, their corresponding mutated MUTs from smallest to largest and stopped when we found a handful of examples that illustrated oracle diversity.
Figure~\ref{fig:rq4} shows two examples from the joda-time and Apache commons-angle package. In the first example, \texttt{test327} calls the \texttt{calculatePrintedLength} method which was mutated by `replacing long division with multiplication', \technique checks the return variable \texttt{int0} and is able to kill the mutant. EvoSuite checks a value that is independent of the execution of the MUT and has no chance to kill the mutant.
The second example, \technique check that \texttt{angle\_Rad1} should not be null and thus was able to detect the mutant which replaces the return value with null. Here again the
EvoSuite generated assertion checks a value that is independent of the MUT.  
While anecdotal these examples illustrate how \technique leverages the EvoSuite generated test prefixes and produces valuable test oracles that EvoSuite can not. 

\begin{tcolorbox}

\textbf{RQ4 Finding:} 
\technique generated test oracles kill 17\% fewer mutants than EvoSuite tests, despite having less information about test behavior.  Relative to TOGA, \technique yields nearly a 10-fold increase in mutant kills - thereby establishing a new SOTA in neural oracle generation.
\end{tcolorbox}

\mr{
\subsection{\textbf{RQ5: Assessing Real Bug Detection Effectiveness}}\label{defects4j} In RQ4, we evaluate \technique's bug detection effectiveness using mutants. In this research question, we investigate \technique's real bug detection effectiveness and compare its performance with TOGA

\subsubsection{Experimental Setup}


To conduct this study, we utilize Defects4J\cite{just2014defects4j}, a benchmark dataset consisting of real Java bugs. To ensure a fair comparison with TOGA, we follow the exact same experimental setup: the same Docker container environment, the same input dataset, and the exact same set of scripts provided in the TOGA replication package\cite{toga-replication-package}. The Defects4J dataset consists of 374 input samples, each with a MUT, a test prefix, and Javadoc documentation. Of these, 304 require an assertion and 70 require an exception oracle. For this dataset, we generate oracle predictions using our method, \technique. Next, \technique generated test oracles are integrated with the respective test prefixes. These integrated test cases (test prefix + oracle) are then executed on both the buggy and fixed program versions. A bug is considered detected if a test passes on the fixed version but fails on the buggy version, indicating that the test oracle correctly aligns with the expected behavior and is capable of detecting the buggy behavior. To understand the results better, we first present the oracle classification performance of both methods in Table \ref{tab:rq5-1}, followed by their bug detection performance in Table \ref{tab:rq5-2}.

\begin{table}[t]
\small\centering
\caption{\mr{Comparison of oracle classification performance between \technique and TOGA on the Defects4J dataset. Metrics include Precision (P), Recall (R), F1-Score (F1), and Accuracy (Accu.).}}\label{tab:rq5-1}
\resizebox{\columnwidth}{!}{%

\begin{tabular}[t]{l|c|c|c|c|c|c|c|c|c}
  \toprule
\multirow{2}{*}  \textbf{Approach} & \multicolumn{3}{c|}{\textbf{Exception}} & \multicolumn{3}{c|}{\textbf{Assertion}} & \multicolumn{3}{c}{\textbf{Overall}}\\
\cline{2-4}  \cline{5-7}  \cline{8-10}
& \textbf{\thead{P}} & \textbf{\thead{R}} & \textbf{\thead{F1}} & \textbf{\thead{P}} & \textbf{\thead{R}} & \textbf{\thead{F1}} & \textbf{\thead{Accu.}} & \textbf{\thead{P}} & \textbf{\thead{R}}\\
\midrule
   
\textbf{TOGA} & .14   & .12   &   .13 & .83  &    .86    & .84 & .74 &  .49 & .49\\
\midrule
\textbf{TOGLL} &  1 &  1 & 1 & 1 & 1 &  1 &  1 &  1 & 1\\	

\bottomrule
\hline
\end{tabular}
}
\end{table}

\subsubsection{Results}Table \ref{tab:rq5-1} presents the oracle classification performance of both methods on the Defects4J dataset. In this classification task, each method determines which type of oracle is needed for a given input prefix. The table shows performance metrics for both classes: exception and assertion, as well as the overall metrics.

TOGA exhibits very poor precision, recall, and F1-score for the class of test prefixes requiring an exception oracle. In contrast, our method, \technique, demonstrates a very high classification accuracy. In RQ2, the exception oracle classification accuracy averaged 93.4\% for 222k input samples. In this smaller study of 374 inputs it achieved an accuracy of 100\%.

Regarding assertion oracles, although TOGA correctly classified that an assertion oracle is needed for 84\% of the inputs, it generated an explicit assertion for only 40\% of them. For 60\% of inputs, TOGA could not generate any assertions. On the other hand, \technique correctly classified that an assertion oracle is needed with 100\% accuracy for the
Defects4J study, and for 97\% of them, it generated an assertion oracle.

Table \ref{tab:rq5-2} shows the bug detection performance of both approaches. Column 2 shows the total number of bugs detected by exception oracles. \technique detected a total of 27 bugs with its exception oracles, whereas TOGA detected only 5. These findings align with the data in Table \ref{tab:rq5-1}, which shows that TOGA's exception oracle generation precision is only 0.14. These results are also consistent with the RQ2 findings, which show a significant performance difference between TOGA and TOGLL, indicating a similar trend across multiple datasets.

Column 3 shows the total number of bugs detected by explicit assertion oracles (non-empty assertions). For the input test prefixes requiring an assertion oracle, TOGA generated assertions for only 40\% of the inputs (121 out of 304), detecting a total of 26 bugs. In contrast, \technique generated assertions for 97\% of the inputs, detecting a total of 37 bugs. 

Column 4 shows the number of bugs detected by test prefixes alone. TOGA could not generate assertions for 183 out of 304 prefixes.  Simply running the test prefixes on the buggy versions detected 36 bugs -- 
these are detected by the implicit oracles of the Java Runtime System that expects no uncaught exception. \technique did not generate an assertion for only 9 prefixes, with only 1 bug detected by running those prefixes. 

As suggested by recent prior work~\cite{10.1145/3611643.3616265, liu2023towards},  bugs detected by implicit oracles when executing test prefixes throwing uncaught exceptions should \textit{not} be considered as contributions of a test oracle generation method. We followed this guidance in RQ4 and applying it here by excluding bugs found in this way
reveals that \technique detects 64 bugs, which is 
106.5\% more bugs than the 31 bugs detected by TOGA. 
Even if one were to count the bugs detected by implicit oracles, \technique detects 16\% more bugs than TOGA on Defects4J.

\begin{table}[t]
\small\centering
\caption{\mr{Defects4J bug detection performance of \technique vs. TOGA. A bug can be detected by different type of oracles.}}\label{tab:rq5-2}
\resizebox{\linewidth}{!}{%

\begin{tabular}[t]{l|c|c|c|c|c}
  \toprule
\multirow{2}{*}  \thead{Approach} & \multicolumn{4}{c|}{Bug Detected By} & FP\\
\cline{2-5}
& \thead{Exception\\ Oracle} & \thead{Explicit \\Assertion \\Oracle} & \thead{EvoSuite \\Test Prefix \\Only} & \thead{Total\\Unique}\\
\midrule
   
\textbf{TOGA} &  5 & 26 &  36 &  56  & 0.45 \\
\midrule
\textbf{TOGLL} & 27 & 37 & 1 & 65 & 0.42 \\
	
\bottomrule
\hline
\end{tabular}
}
\end{table}

\begin{tcolorbox}

\textbf{RQ5 Finding:} 
\mr{When excluding the bugs that are detected by test prefixes alone, TOGA detected a total of 31 Defects4J bugs with its exception and explicit assertion oracles. In comparison, \technique detected 64 bugs and improvement of 106\%.}
\end{tcolorbox}
}

\subsection{Threats to Validity}
In our large-scale study, we have utilized open-source and widely recognized SF110 dataset \cite{10.1145/2685612}. Yet, our results may not generalize across other datasets. \mr{To ensure generalizability, we further evaluated \technique on the unseen OracleEval25 ~\cite{10.1145/3611643.3616265} and Defects4J \cite{just2014defects4j} dataset.}

In our comprehensive study, we created several tools and scripts which may contain bugs.  We used publicly available libraries to mitigate the risk and conducted extensive validity tests and repeated our experiments to ensure consistent results. 

In our investigation, during fine-tuning we utilized the `exact match'  metric to compute oracle generation accuracy. This is a widely used metric by the research community. For unseen inference, we relied on test validation. This methodology, we assert, offers a reliable accuracy metric. \mr{Additionally, to complement the results from mutation testing, we performed a real bug detection study. Both metrics are well-established and used in many research studies~\cite{zhang2015assertions,10.1145/3611643.3616265,hossain2023measuring}.}


\section{Limitation And Future Direction}
In this work, we have demonstrated the potential of LLMs in generating correct, diverse and strong test oracles that can effectively detect large number of unique bugs. However, there are challenges  which require further research efforts.

\subsection{Compilation Error} Our in-depth experimental results show that \technique generates assertions that are non-compiling about 5\% of the time. Among them we have seen assertions that are `garbage' strings and one's that have minor syntax or type errors.  In future work, we want to fine-tune LLMs to generate syntax and type correct assertion using grammar and type based constraints similar to \cite{10.1145/3510003.3510141}. Additionally, incorporating compilation errors into a refinement prompt to improve the the generated oracle is another strategy worth pursuing~\cite{deligiannis2023fixing}.

\subsection{Token Length Limits} We have seen that LLMs can not process input samples that exceeds the maximum token length capacity of the LLMs. In our study, a total of 3\% of the samples exceeded the threshold of 600 tokens when using P6 that includes MUT, docstring and prefix. 
Rather than increasing the token threshold, our study suggests that in such cases we could shift to a more concise prompt, such as P3 which replaces the MUT with the method signature without significantly compromising performance. 

\subsection{False Positives} One important and long standing issue with all type of learning based oracle generation method is that they can generate false positives, i.e., oracles that fail on a correct program. 
\technique exhibited a 7\% false positive rate for exception oracles
and a 
25\% rate for assertion oracles which is a significant improvement over the respective rates for TOGA -- 81\% and 47\%. 
We have studied a sample of false positive assertions and observed that many of them
include numeric literals which is the source of the false positive.  Based on these
observations, we conjecture that
assertions relating program variables, or expressions involving variables,
may be less susceptible to false positives.  
In future research, we explore the benefits of using high quality documentation, which may capture
such relations, and develop prompting strategies that encourage the generation of 
assertions that are not as dependent on numeric literals.

\section{Related Work}
EvoSuite~\cite{fraser2011evosuite} is a state of the practice automated test oracle generation that uses a search-based method. EvoSuite generates oracles by observing test execution and capturing values that it includes in assertions.  This creates  regression oracles that are valuable for detecting undesirable changes in program behavior.  Importantly for our study, by construction these oracles capture the behavior of the system under test making them a useful baseline for judging correct test oracles.

Randoop~\cite{pacheco2007randoop} is feedback-directed random testing for Java. 
Research has explored the generation of both assertion~\cite{10.1145/3213846.3213872} and exception 
oracles~\cite{10.1145/2931037.2931061} for Randoop tests using NLP approaches.
Subsequent NLP-based methods, such as ATLAS~\cite{watson2020learning} and AthenaTest~\cite{tufano2020unit}, have been shown to outperform these methods and
 more recently TOGA ~\cite{10.1145/3510003.3510141} established a new state
of the art by a large margin.  This is why we use TOGA as baseline in our study. 

In the past year, techniques for test generation using LLMs have emerged. For example, TESTPILOT ~\cite{10329992}, an LLM-based test generation tool for JavaScript that automatically generates unit tests for JavaScript. TESTPILOT achieved 70\% statement coverage and 52.8\% branch coverage, which outperformed the state-of-the feedback-directed JavaScript
test generation technique, Nessie~\cite{arteca2022nessie}. A recent study ~\cite{siddiq2024using} that utilized LLMs for generating test for the SF110 large-scale real world Java projects showed that LLMa could only achieve 2\% coverage, whereas EvoSuite can achieve more than 90\% coverage.  This suggests that LLMs are not very effective in generating test for real world Java program, which is why in our study instead of focusing on both test input and test oracle generation, we focus on a single task test oracle generation with LLM. 


\section{Conclusion} 
\mr{We conduct the first large-scale investigation of the ability of large language models to automatically generating correct and strong test oracles. To this end, we fine-tune seven code LLMs with six different prompts with varying level of information. Our findings reveal that effective prompt design can improve the accuracy of the models significantly and that with effective prompts smaller size model can also perform equally or even better than larger models. 

With the best performing model and prompt we develop \technique, our LLM-based automated test oracle generation method. We extensively evaluate \technique through large-scale study to investigate its generalizability and capability to generate correct, diverse and strong test oracles. Our experimental study answers critical research questions, demonstrating that \technique can generate 3.8x more correct assertion and 4.9x more exception oracles than the previous SOTA neural method, TOGA. 
Generating correct oracle does not necessarily mean they are diverse and strong enough to detect bugs~\cite{10.1145/3639478.3639791}. We find that \technique is capable of generating diverse test oracle where only 9.5\% are an exact match with the ground truth. Furthermore, \technique's diverse assertions can detect a large number of unique bugs  that EvoSuite can not detect. For both mutant and real bug detection, \technique significantly outperformed TOGA, establishing itself as the new SOTA. 

We describe remaining challenges and discuss several actionable future directions. We believe our large-scale study and in-depth investigation lay down the groundwork
for further advancements in automated test oracle generation.}

\section{Acknowledgment}
This material is based in part upon work supported by National Science Foundation awards 2129824 and  221707.  The authors acknowledge Research Computing at The University of Virginia for providing computational resources and technical support that have contributed to the results reported within this publication.
\bibliographystyle{IEEEtran}
\bibliography{ICSE-25}
\end{document}